\documentclass{aa}
\usepackage{amsmath}
\usepackage[dvips]{graphicx}
\usepackage{psfig}
\usepackage{times}

\def\mabs{$M_{\rm B}$}

\def\hii{H{\sc ii}}

\def\telec{$T_{\rm e}$}

\def\tnii{$t_{\rm [N\, II]}$}
\def\doh{$12 + \log(\rm O/H)$}

\def\micron{$\mu$m}
\def\kms{km s$^{-1}$}
\def\kmsmpc{km s$^{-1}$ Mpc$^{-1}$}

\def\ergscm{ergs s$^{-1}$ cm$^{-2}$}

\def\msun{M$_{\odot}$}
\def\msunyr{M$_{\odot}$ yr$^{-1}$}
\def\zsun{Z$_{\odot}$}
\def\halpha{\ifmmode {\rm H{\alpha}} \else $\rm H{\alpha}$\fi}
\def\hbeta{\ifmmode {\rm H{\beta}} \else $\rm H{\beta}$\fi}
\def\oii{[O\,{\sc ii}] $\lambda\lambda$3726,3728}
\def\oiia{[O\,{\sc ii}] $\lambda$3726}
\def\oiib{[O\,{\sc ii}] $\lambda$3728}
\def\oiii{[O\,{\sc iii}] $\lambda\lambda$4959,5007}
\def\oiiia{[O\,{\sc iii}] $\lambda$4959}
\def\oiiib{[O\,{\sc iii}] $\lambda$5007}
\def\oiiic{[O\,{\sc iii}] $\lambda$4363}
\def\ntwo{[N\,{\sc ii}]}
\def\nii{[N\,{\sc ii}] $\lambda$6584}

\def\niisha{[N\,{\sc ii}]/H$\alpha$}
\def\niisoii{[N\,{\sc ii}]/O\,{\sc ii}]}

\def\rr23{$R_{\rm 23}$}
\def\oo32{$O_{\rm 32}$}
\def\xx3{$X^{*}_{\rm 3}$}
\newcommand{\ltapprox}{\raisebox{-0.5ex}{$\,\stackrel{<}{\scriptstyle\sim}\,$}}
\newcommand{\gtapprox}{\raisebox{-0.5ex}{$\,\stackrel{>}{\scriptstyle\sim}\,$}}

\begin{document}

\title{Physical properties of two low-luminosity $z \sim 1.9$ galaxies behind 
the lensing cluster AC 114\thanks{Based on observations collected at the Very Large Telescope (Antu/UT1), 
European Southern Observatory, Paranal, Chile (ESO Programs 64.O-0439, 67.A-0466)}}

\author{M. Lemoine-Busserolle\inst{1}, T. Contini\inst{1},
R. Pell\'o\inst{1}, J.-F. Le Borgne\inst{1}, J.-P. Kneib\inst{1,2}, C. Lidman\inst{3}}

\titlerunning{Physical properties of two low-luminosity $z \sim 1.9$ galaxies}

\authorrunning{M. Lemoine-Busserolle et al.}

\offprints{M. Lemoine-Busserolle; {\tt marie.lemoine@ast.obs-mip.fr}}

\institute{Laboratoire d'Astrophysique de l'Observatoire Midi-Pyr\'en\'ees - UMR 5572,
14 Avenue E. Belin, F-31400 Toulouse, France
\and
California Institute of Technology - Pasadena, CA 91125, USA
\and
European Southern Observatory - Alonso de Cordova 3107, Vitacura, Chile
}

\date{Received 2002 September 4; accepted 2002 October 24}

\abstract{We present VLT/ISAAC near-infrared spectroscopy of two
gravitationally-lensed $z \sim 1.9$ galaxies, A2 and S2, located behind
the cluster AC 114. Thanks to large magnification factors, we have
been successful in detecting rest-frame optical emission lines (from
[O\,{\sc ii}]$\lambda$3727 to \halpha+[N\,{\sc ii}]$\lambda$6584) in
star-forming galaxies 1 to 2 magnitudes fainter than in previous
studies of Lyman break galaxies (LBGs) at $z \sim 3$.  From the
\halpha\ luminosity, we estimate star formation rates (SFRs) of 30 and
15 \msunyr\ for S2 and A2 respectively. These values are 7 to 15 times
higher than those inferred from the UV continuum flux at 1500\AA\
without dust extinction correction. In setting SFR$_{\rm H\alpha}$
$\sim$ SFR$_{\rm UV}$, one derives extinction coefficients 
E(B-V) $\sim$ 0.3 for S2 and E(B-V) $\sim$ 0.4 for A2.  
The behavior of S2 and A2 in terms of O/H and N/O abundance ratios 
are very different, and they are also
different from typical LBGs at $z \sim 3$. S2 is a low-metallicity
object ($Z \sim$ 0.03 \zsun) with a low N/O ratio, similar to those
derived in the most metal-poor nearby \hii\ galaxies. In contrast, A2
is a high-metallicity galaxy ($Z \sim$ 1.3 \zsun) with a high N/O abundance
ratio, similar to those derived in the most metal-rich starburst nucleus 
galaxies. The line-of-sight velocity dispersions, derived from emission 
line widths, are 55 and 105 \kms, yielding a virial mass of 0.5 and 
2.4 $\times 10^{10}$ \msun, for S2 and A2 respectively. 
Thanks to the gravitational amplification, the line profiles of S2 are 
spatially resolved, leading to a velocity gradient of $\pm 240$ \kms, 
which yields a dynamical mass of $\sim 1.3 \times 10^{10}$ \msun\ within 
the inner 1 kpc radius.
Combining these new data with the sample of LBGs at $z \sim 3$, including 
the lensed galaxy MS 1512-cB58, which is the only LBG for which physical 
properties have been determined with similar accuracy, we conclude that 
these three galaxies exhibit different physical properties in terms of 
abundance ratios, SFRs, $M/L_{\rm B}$ and reddening. High-redshift 
galaxies of different luminosities could thus have quite different 
star formation histories.}

\maketitle

\keywords{galaxies: evolution -- galaxies: starburst -- galaxies:
  abundances -- galaxies: kinematics -- infrared: galaxies}

\section{Introduction}

With the recent advent of near-infrared (NIR) spectrographs on 8-10m
class telescopes, a quantitative study of the rest-frame optical
properties of high-redshift galaxies is now possible. Indeed, NIR
spectroscopy allows one to compare directly the physical properties of
high-redshift galaxies (star formation rate, reddening, metallicity,
kinematics, virial mass) with galaxies in the local Universe through
line-based indicators.

The pioneering work by Pettini et al. (1998, 2001) has shown
that the rest-frame optical properties of the brightest Lyman break
galaxies (LBGs) at $z \sim 3$ are relatively uniform.  
However, the sample of LBGs observed in the NIR is small and limited to
the brightest examples (Pettini et al. 1998, 2001; Kobulnicky \& Koo
2000; Teplitz et al. 2000; Shapley et al. 2001).  Additionally, the
$1.5 \ltapprox z \ltapprox 2.5$ redshift interval remains unexplored
because of the lack of strong spectral features that can be used to
identify such sources with visible spectrographs. 

Massive clusters acting as Gravitational Telescopes (GTs) constitute a
powerful tool in the study of high-redshift galaxies. They have been
successfully used over a wide range of wavelengths, ranging from the UV 
to the sub-mm (e.g. Smail et al. 1997; B\'ezecourt et al. 1999; Ebbels et
al. 1998; Pell\'o et al. 1999; Altieri et al. 1999; Ivison et
al. 2000) and allowed recently to detect the most distant galaxies 
($z \ga 5$; Ellis et al. 2001; Hu et al. 2002). 
The large magnification factors of galaxies that are 
close to the critical lines (typically 1 to 3 magnitudes) 
can be used to probe the physical properties of intrinsically faint
high-redshift galaxies, which would otherwise be beyond the
limits of conventional spectroscopy (e.g. Ebbels et al. 1996; Pell\'o
et al. 1999, Mehlert et al.  2001). 

Large samples of $z \gtapprox 2.5 $ star-forming galaxies have become
available in the recent years, mainly through the Lyman break
technique (Steidel et al. 1996, 1999). Our sample of highly-magnified,
high-redshift galaxies extends these samples towards {\it
intrinsically fainter galaxies}, allowing us to probe the physical
properties of galaxies that are 1 to 3 magnitudes fainter than the
present LBGs studies (Steidel et al. 1996, 1999; Pettini et al. 1998,
2001). 
The sample has been selected from lensing clusters with well
constrained mass distributions, first through photometric
redshift and lens-inversion techniques, and later confirmed
spectroscopically (e.g. Kneib et al. 1996; Ebbels et al. 1998;
Campusano et al. 2001). Lensed galaxies for spectroscopic follow-up
are chosen to be close to the high-redshift critical lines in order to
obtain the largest magnification (typically $\sim 2$ magnitudes).
 
In this paper, we present NIR spectroscopy on two high-redshift
galaxies ($z \sim 1.9$) that are lensed by the massive cluster AC 114
($z=0.312$). AC 114 is a rich source of multiply-imaged, high-redshift
galaxies (Smail et al. 1995; Natarajan et al. 1998, Campusano et
al. 2001). The two lensed galaxies, named S2 and A2 according to
Natarajan et al. 1998, correspond to multiple images of two different
sources at very similar redshifts.  A mass-model, first obtained by
Natarajan et al. (1998) and later improved by Campusano et al. (2001), 
allows us to recover the intrinsic properties of S2 and A2.  Thanks to
the large magnification of these sources, we have measured for the
first time several important emission lines from NIR
spectroscopy of high-redshift galaxies fainter than \mabs\ = $-21$.

The plan of the paper is as follows. In Sections \ref{observations}
and \ref{data}, we summarize the observations and describe the steps
in converting the two dimensional raw data into calibrated one
dimensional spectra. Emission-line intensities and FWHMs are presented in
Section~\ref{results} and analysed in Section~\ref{analysis}.  Star
formation rates, O/H and N/O abundance ratios and kinematics are
derived here.  In Section~\ref{discussion} these new data are
combined with existing samples of LBGs at $z \sim 3$ in order to
understand where high-redshift galaxies lie in two fundamental scaling
relations: N/O versus O/H abundance ratios and the metallicity--luminosity
relation. Finally, the main results of this work are summarized in
Section~\ref{conclusions}. 
Throughout this paper, we assume a cosmology with $\Omega_{0} = 0.3$,
$\Lambda = 0.7$ and $H_{0}$ = 70 \kmsmpc.

\section{Observations}
\label{observations}

The data were obtained with the NIR spectrograph ISAAC, which is
located on the Nasmyth B focus of Antu (VLT-UT1), during UT 2001
September 26-27 (period 67). The data were obtained with the
Short-wavelength channel (Cuby et
al. 1999), which uses a $1024 \times 1024$ Hawaii Rockwell array, and
the medium resolution grating. The slit width was 1{\arcsec}.

Because the redshifts and spectral energy distributions (hereafter
SED) of the sources in the cluster core were known before the run, we
could optimize the use of the grating settings to search for the most
relevant spectral features. To cover \halpha, \hbeta, \oiii, and \oii,
three settings with the ISAAC medium resolution grating were used.

\begin{figure}[t]
\centering
\includegraphics[width=9.0cm]{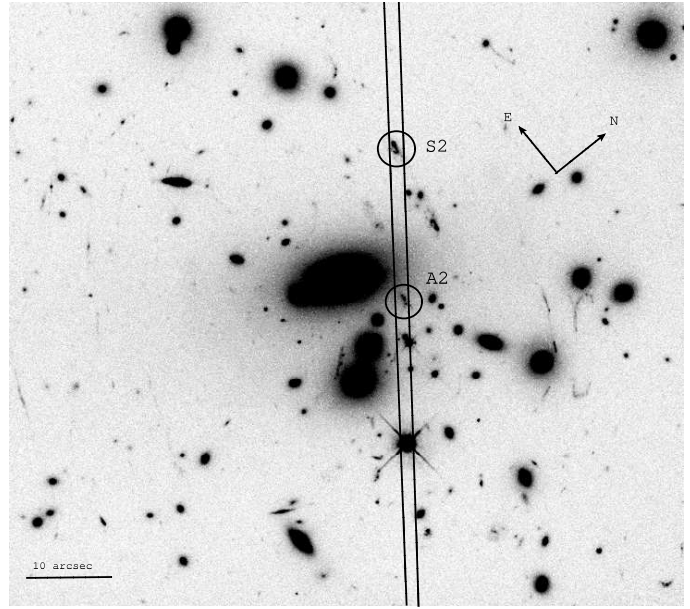}
\caption{HST/WFPC2 image ($R$-band/F702W) of the core of AC114.  The
position of the ISAAC slit used in our
observations is displayed. Circles mark the position of the high redshift galaxies
AC114-S2 and AC114-A2.}
\label{ima}
\end{figure}

The ISAAC 120{\arcsec} long slit allowed us to align at least 2
objects per setting in the inner region of the cluster.  In the
present case, we targeted the two gravitationally amplified sources
AC114-S2 and AC114-A2, both of them at redshift $z \sim 1.9$, together
with the offset star. Figure~\ref{ima} shows an HST/WFPC2 image
(F702W) of AC114 and the long-slit configuration. 

Table~\ref{tab1} summarizes the wavelength regions that were selected
to contain the redshifted emission lines of \oii, \oiii, \halpha, and
\nii. The integration times and the airmasses are also listed.

\begin{table}[t]
\begin{center}
\caption{Summary of ISAAC Observations}
\label{tab1}
	\begin{tabular}{l c c c}
	\hline \hline
	& [OII] & [OIII]  & \halpha\ \\
	\hline 
	Atm. window& $Y$  & $H$  & $K$ \\
	$\lambda$ (\AA) observed & 10500-10840  & 13900-14500  & 18400-19400 \\
	$\lambda$ (\AA) rest-frame & 3662-3780  & 4848-5057 & 6417-6766 \\
	Exposure (s) & $8 \times 900$  & $8 \times 900$ & $8 \times 900$ \\
	\hline 	    
	\end{tabular}	
\end{center}
\end{table}

\begin{figure*}[t]
\centering
\begin{center}
$Y$-band \hspace{5cm} $K$-band
\end{center}
\vspace{-0.5cm}
\includegraphics[height=4.8cm]{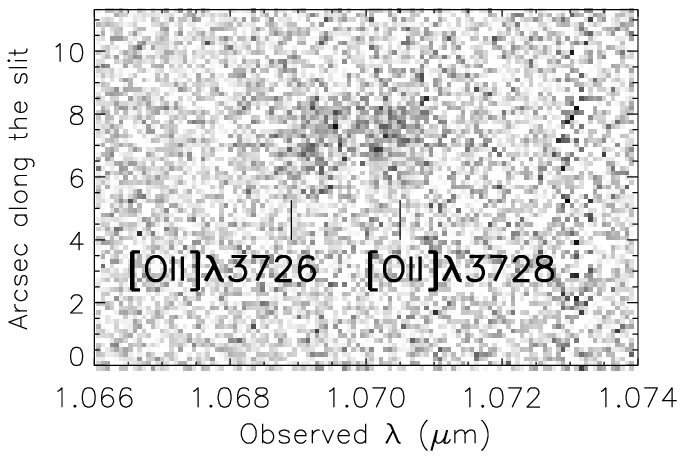}
\includegraphics[height=4.8cm]{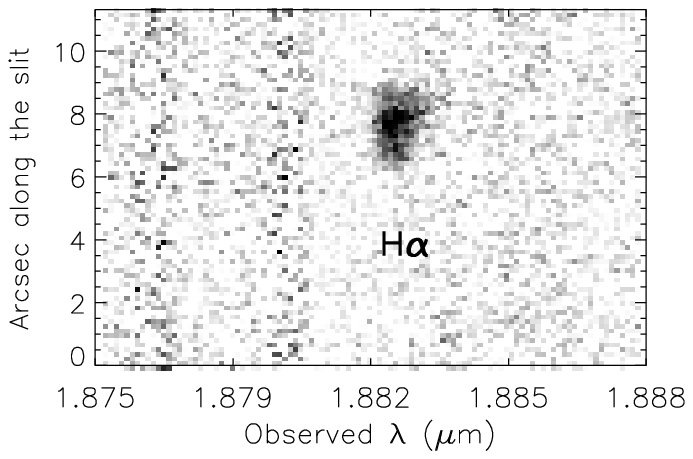}\\
\begin{center}
$H$-band 
\end{center}
\vspace{-0.5cm}
\includegraphics[height=4.8cm]{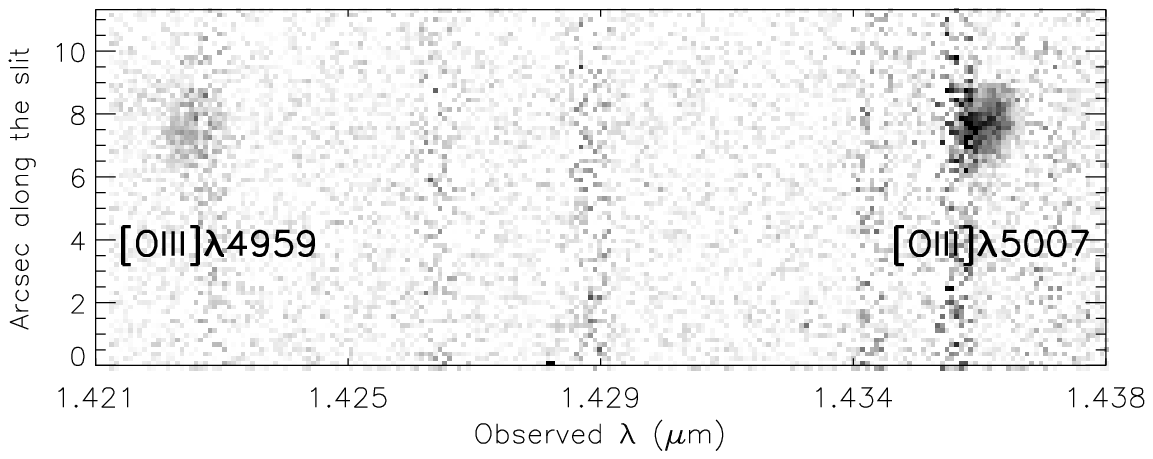}\\
\caption{Two-dimensional images of the ISAAC spectra of AC114-S2 in the 
$Y$ (1.0 to 1.1 \micron), $H$, and $K$ bandpasses. Positions of the nebular emission lines 
are labeled.}
\label{ima2d}
\end{figure*}

The ISAAC observations were performed in beam-switching mode. The
objects were observed at two slit positions, which we label A and
B. After an A-B-B-A sequence, the object was re-acquired at a different
slit position, and the sequence was repeated. The seeing varied 
between $\sim 0.5$ and $\sim 1.5$ arcsec and airmass varied between 
1.2 and 1.4.

\begin{figure}[t]
\centering
\includegraphics[width=8.5cm]{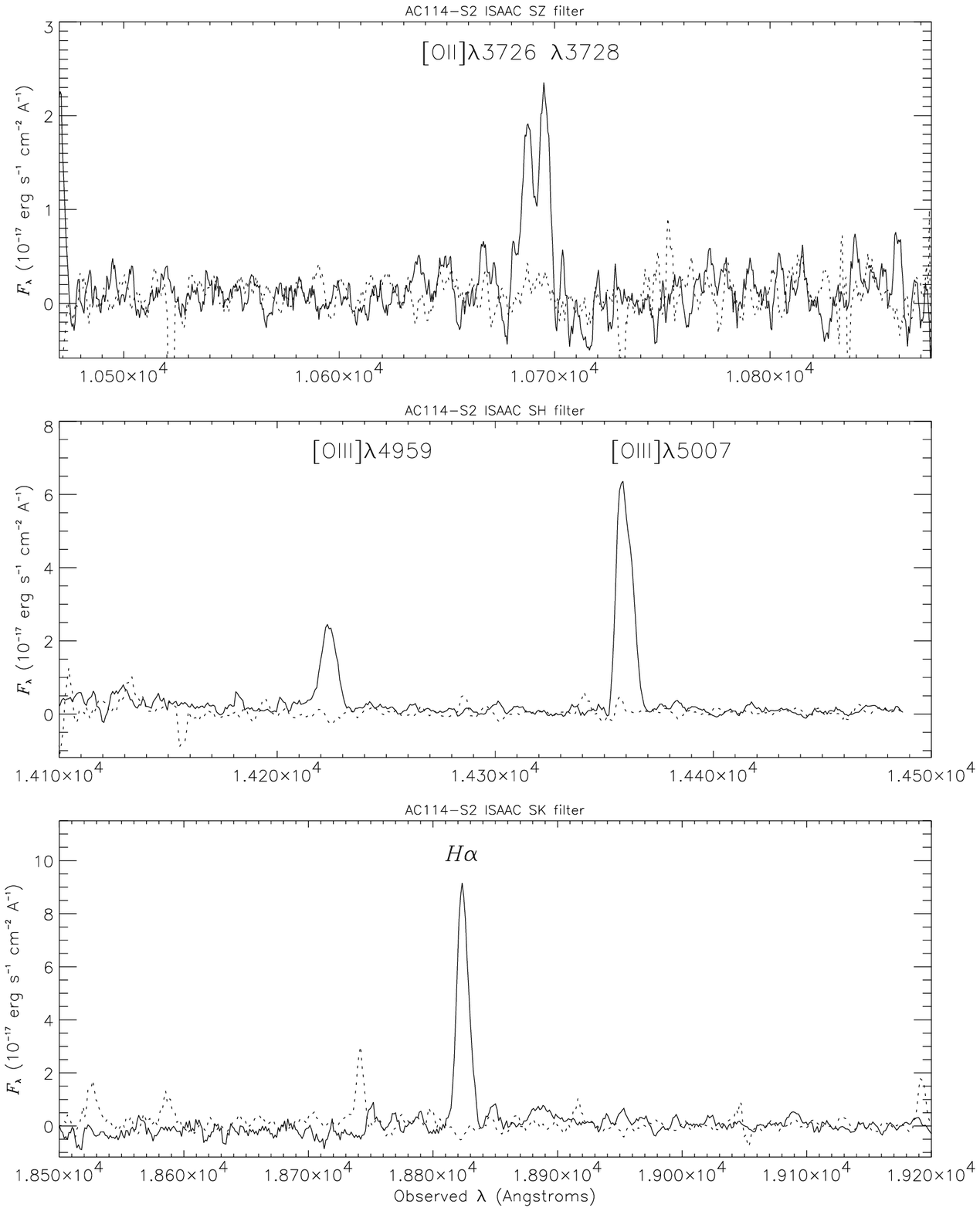}\\
\caption{The extracted one-dimensional $Y$, $H$, and $K$ band 
spectra of AC114$-$S2. The dashed line shows the 1 $\sigma$ error spectrum. 
Spectra have been smoothed to the instrumental resolution.
}
\label{ima1ds2}
\end{figure}

\begin{figure}[t]
\centering
\includegraphics[width=8.5cm]{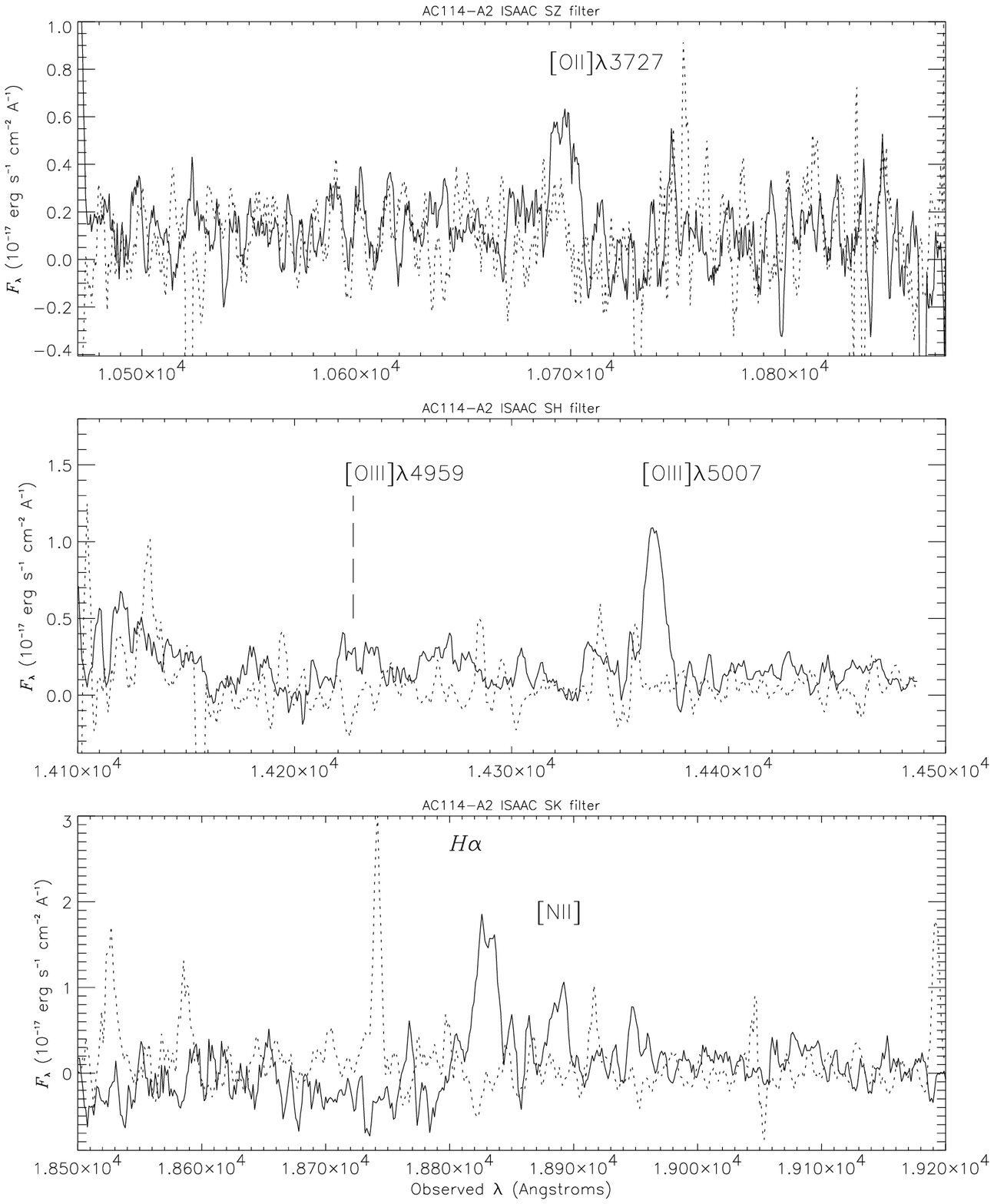}\\
\caption{The extracted one-dimensional $Y$, $H$, and $K$ band 
spectra of AC114$-$A2. The dashed line shows the 1 $\sigma$ error spectrum. 
Spectra have been smoothed to the instrumental resolution.
}
\label{ima1da2}
\end{figure}

\section{Data reduction and calibration}
\label{data}

We used IRAF with reference to the ISAAC Data Reduction Guide
\footnote{ISAAC Data Reduction Guide: 
{\tt http://www.hq.eso.org/instruments/isaac/index.html}} to convert the
two-dimensional raw data into calibrated one-dimensional spectra.  The
data are first split into AB pairs, and the first step in the
reduction process (the first sky subtraction) is to subtract one
frame from another in each pair.  This step removes both the zero
level offset of the array and the sky emission, and results in images
with two spectra, one positive and one negative. After trimming and
flat fielding, we applied a two-dimensional wavelength calibration.
The wavelength calibration, which was derived from the OH lines
(Rousselot et al. 2000), also corrects for slit curvature. The next
step, the second sky subtraction, removes any residual sky features
remaining after the first sky subtraction. During this process, each
image is multiplied by $-1$ and added back to itself after a suitable
shift.  It results in images that have one positive spectrum and two
negative spectra either side of the positive spectrum. The resulting
spectra, one for each AB pair, are then combined. As an example, the
two dimensional spectrum of S2 is shown in Figure~\ref{ima2d}.  The
emission lines \oii, \oiii, and \halpha\ are all clearly detected.

We extracted a one dimensional spectrum from two-dimensional data
without fitting the sky since it would only add noise. Removing
telluric lines is done by dividing the object spectrum with that of a
telluric standard. The telluric standard was observed with the same
instrument setup, immediately after the science targets, and it was
reduced in the same way and with the same calibration frames as for
the science targets. Since the telluric standards were hot stars, a
blackbody curve was used to model the continuum of these
standards. The blackbody temperature was fixed according to the
spectral type of the star. 
The accuracy of telluric line
removals has been checked by computing the observed \oiiib/\oiiia\
line ratio (see Table~\ref{tab2}).  For S2, we find \oiiib/\oiiia\ =
$2.5 \pm 0.4$ which, given the measurement uncertainties, agrees with
theoretical value of 2.86 (e.g. Osterbrock 1989).
Figures~\ref{ima1ds2} and~\ref{ima1da2} show the extracted
one-dimensional infrared $Y$, $H$, and $K$ band spectra of S2 and A2
respectively.

\section{Emission line measurements}
\label{results}

The emission lines were fitted with one dimensional Gaussians. The
integrated fluxes (or 2$\sigma$ upper limits when appropriate) and velocity
widths are listed in Table~\ref{tab2}.
The intrinsic velocity width of emission lines was corrected
for the instrumental profile, which is determined from the width
of night-sky lines. The redshifts measured for S2 and A2 are 1.867 and 1.869
respectively. In the case of S2, this value agrees with
previous measurements (Smail et al. 1995; Campusano et al. 2001). 
In the case of A2, there is disagreement between the present secure 
value and the estimate by Campusano et al. (2001), $z=1.687$.
Taking into account the correction due to gravitational lensing, the 
distance between the two objects is about 4\arcsec\ on the source 
plane at $z=1.869$, corresponding to $\sim 34$ kpc with the adopted 
cosmology. As they have similar redshifts, S2 and A2 are thus very 
close in the source plane and are probably associated with one another.
Given the redshifts of A2 and S2, \hbeta\ is not detected because it
is heavily obscured by telluric absorption between the $Y$ and $H$
bands.

The lensing model by Campusano et al. (2001) was revised to take into
account the correct value for the redshift of A2.  The resulting
magnification factors, 2.0 magnitudes for S2 and 1.7 magnitudes for
A2, are similar to the previous estimates (Campusano et al. 2001) 
and within the model errors ($\sim 0.1$ magnitudes).

\begin{table*}[t]
\begin{center}
\caption{Spectral Measurements: Line fluxes and FWHM}
\label{tab2}
	\begin{tabular}{l c c c }
	\hline \hline
	Emission Line & AC114-S2 & AC114-A2 & Comments\\
	\hline 
	\oiia\ :  					  & 		   & & \\
	\hspace{0.3cm} $F(\times 10^{-17}$ \ergscm) & $1.8\pm 0.5$ & &	(1)\\
	\hspace{0.3cm} FWHM (\AA)  		          & $5.4\pm 0.7$ & & (3)\\
	\hspace{0.3cm} $\sigma$ = FWHM/2.35 (\kms) 	  & $65\pm 8$& &	(3)\\
	\oiib\ :  					  & 		   & \oii\ : &	\\
	\hspace{0.3cm} $F(\times 10^{-17}$ \ergscm) & $1.9\pm 0.4$ & $1.5\pm 0.4$ &	(1)\\
	\hspace{0.3cm} FWHM (\AA) 	  		  & $4.2\pm 0.4$ & $11.9\pm 1.8$&	(3)\\
	\hspace{0.3cm} $\sigma$ = FWHM/2.35 (\kms)  	  & $50\pm 5$& $142\pm 22$&(3) \\
	\hline
	\oiiia\ : 					  & 		   & &	\\
	\hspace{0.3cm} $F(\times 10^{-17}$ \ergscm) & $3.4\pm 0.4$ & $< 0.9$ (2)&	(1)\\
	\hspace{0.3cm} FWHM (\AA) 	  		  & $6.9\pm 0.5$ &  ...&	(4)\\
	\hspace{0.3cm} $\sigma$ = FWHM/2.35 (\kms) 	  & $62\pm 5$&  ...&	(4)\\
	\hline	
	\oiiib\ : 					  & 		   & &	\\
	\hspace{0.3cm} $F(\times 10^{-17}$ \ergscm) & $8.6\pm 0.4$ & $2.5\pm 0.4$&	(1)\\
	\hspace{0.3cm} FWHM (\AA) 	  		  & $6.4\pm 0.2$ & $9.5\pm 0.4$&	(4)\\
	\hspace{0.3cm} $\sigma$ = FWHM/2.35 (\kms) 	  & $57\pm 2$& $84\pm 3$&	(4)\\
	\hline
	\halpha 				  & 		   & &	\\
	\hspace{0.3cm} $F(\times 10^{-17}$ \ergscm) & $15.6\pm 0.9$& $7.7\pm 0.8$ &(1)	\\
	\hspace{0.3cm} FWHM (\AA) 	  		  & $6.6\pm 0.5$ & $18.5\pm 0.9$ &	(5)\\
	\hspace{0.3cm} $\sigma$ = FWHM/2.35 (\kms)  	  & $45\pm 3$& $125\pm 6$&(5) 	\\
	\hline
	\nii\ :					  & 		   & &	\\
	\hspace{0.3cm} $F(\times 10^{-17}$ \ergscm) & $< 0.8$ (6)	   & $3.3\pm 0.4$ &	(1)\\
	\hspace{0.3cm} FWHM (\AA) 	  		  & ...            & $13.6\pm 0.9$ &	(5)\\
	\hspace{0.3cm} $\sigma$ = FWHM/2.35  (\kms)  	  & ...            & $92\pm 6$&(5) \\
	\hline 	 \\   
	\end{tabular}
\end{center}
COMMENTS: (1) Measurement by Gaussian fit, corrected for the gravitational 
magnification factor of 6.31 for AC114-S2 and 4.79 for AC114-A2. (2) Assuming 
$F_{5007}/2.9$. (3) Intrinsic FWHM after correction for the instrumental profile, 
modeled as a Gaussian with FWHM=2.3\AA. (4) Intrinsic FWHM after correction for 
the instrumental profile, modeled as a Gaussian with FWHM=4.3\AA. (5) Intrinsic 
FWHM after correction for the instrumental profile, modeled as a Gaussian with 
FWHM=6.8\AA. (6) $2 \sigma$ upper limit. 
\end{table*}

\section{Analysis}
\label{analysis}

\subsection{Star formation rates}
\label{SFR}

We have compared the star formation rate (SFR) obtained from the UV
continuum flux with the value derived from the \halpha\ emission line,
which is the indicator used for galaxies in the local Universe
(e.g. Gallego et al. 1995). The rest-frame 1500\AA\ flux for S2 and A2
has been derived from the UV rest-frame FORS1 spectra obtained by
Campusano et al. (2001). The spatial sampling of these spectra is
similar to the ISAAC spectra, thus corresponding to the same physical 
region in the two cases.
We have checked these values independently using the best-fit SEDs
derived from $UBVRIJK$ photometry (see details in Campusano et
al. 2001). This step was important for A2, because the FORS1 spectra
have a relatively low S/N ratio.  Star formation rates have been
computed via the following equations (Kennicutt 1998):

\begin{equation}
{\rm SFR_{UV}}(M_{\sun} {\rm yr}^{-1})=1.4\times 10^{-28} L_{1500}({\rm ergs}\ {\rm s}^{-1}\ {\rm Hz}^{-1}),
\end{equation}
\begin{equation}
{\rm SFR_{H\alpha}}(M_{\sun} {\rm yr}^{-1})=7.9\times 10^{-42} L_{\rm H\alpha}({\rm ergs}\ {\rm s}^{-1})
\end{equation}

For our adopted cosmology, the SFRs deduced from the \halpha\ 
luminosities are 30.3 and 15.0 \msunyr\ for AC114-S2 and AC114-A2, 
respectively. These values are 7-15 times higher than those deduced 
from the continuum flux at 1500\AA\ (4.4 \msunyr\ for AC114-S2 and 1.1 
\msunyr\ for AC114-A2) when no correction for extinction 
is applied. If we require that SFR$_{\rm H\alpha}$ = SFR$_{\rm UV}$, then
the extinction coefficients are E(B-V)=0.3 for AC114-S2 and E(B-V)=0.4 
for AC114-A2. With these values, the corrected SFRs become 73 \msunyr\ 
for AC114-S2 and 51 \msunyr\ for AC114-A2, applying the reddening law of 
Calzetti et al. (2000). The reddening value of S2 is in good agreement 
with that derived from the UV rest-frame spectrum (Le Borgne et al., 
in preparation).

These extinctions coefficients are larger than the typical value for 
LBGs up to $z \sim 3$ (Steidel et al. 1999), and correspond to the upper 
envelope of values derived for these objects (Shapley et al. 2001). 
In the remainder of this paper, all line
fluxes, and hence all the quantities that are derived from these
fluxes, will be corrected for extinction.

\subsection{Chemical abundances}
\label{chemabund}

We have calculated oxygen-to-hydrogen (O/H) and nitrogen-to-oxygen
(N/O) abundance ratios for each galaxy using the measured
emission-line fluxes reported in Table~\ref{tab2}. Kobulnicky et
al. (1999) have shown that emission-line ratios integrated over the whole
galaxy can provide a reliable indication of the oxygen and nitrogen
abundances in high-redshift galaxies.
 
\subsubsection{Oxygen abundance}

\begin{table*}[t]
\begin{center}
\caption{Results of the chemical abundance analysis. O/H and N/O abundance ratios using 
different calibrations for the galaxies S2 and A2 in the lensing cluster AC 114. Two cases 
are considered: without extinction correction and assuming a reddening coefficient 
E(B-V)=0.3 and 0.4 for S2 and A2 respectively. Preferred O/H abundance ratios are 
indicated in boldface.}
\label{tab4}
\begin{tabular}{lrrrrl}
\hline
\hline
      &\multicolumn{2}{c}{AC114-S2}&\multicolumn{2}{c}{AC114-A2}& Calibration \\
\hline
E(B-V)&    0.0 & 0.3              & 0.0 & 0.4 &             \\
\hline
\doh\ &        &                  &       &              \\        
lower branch &  {\bf 7.17} & {\bf 7.42}     & 6.88 & 7.25 & from \rr23\ (Pilyugin 2000) \\
lower branch &  {\bf 7.34} & {\bf 7.66}     & 7.16 & 7.56 & from \rr23\ and \oo32\ (Kobulnicky et al. 1999) \\
upper branch &  8.81 & 8.63     & {\bf 8.89} & {\bf 8.68} & from $P$ and \rr23\  (Pilyugin 2001) \\
upper branch &  8.91 & 8.78     & {\bf 8.99} & {\bf 8.86} & from \rr23\ and \oo32\ (Kobulnicky et al. 1999) \\
\multicolumn{6}{l}{\sl Secondary abundance indicators}      \\  
log(\niisha)   & $<-$1.29 & $<-$1.29 & $-$0.36 & $-$0.36 & \\ 
log(\niisoii)  & $<-$0.66 & $<-$0.97 &    0.36 & $-$0.05 & \\	
\hline
log(N/O) & $<-$1.32 & $<-$1.51 & $-$0.42 & $-$0.68 &  from Thurston et al. (1996) \\
\hline 	    
\end{tabular}
\end{center}
\end{table*}

Emission lines are the primary source of information regarding chemical
abundances within \hii\ regions. The ``direct'' method (referred as the 
\telec\ method) for determining 
chemical compositions requires the electron temperature and the density 
of the emitting gas  (e.g., Osterbrock 1989). Unfortunately, a direct 
heavy element abundance determination, based on measurements of the 
electron temperature and density, cannot be obtained for distant galaxies. 
The \oiiic\ auroral line, which is the most commonly applied temperature 
indicator in extragalactic \hii\ regions, is typically very weak and 
rapidly decreases in strength with increasing abundance.

Given the absence of \oiiic\ in our faint spectra, alternative methods
for deriving nebular abundances that rely on the bright lines alone
must be employed.  Empirical methods to derive the oxygen abundance
exploit the relationship between O/H and the intensity of the strong
lines via the parameter \rr23\ $\equiv$
(\oiia+\oiib+\oiiia+\oiiib)/\hbeta\ both for metal-poor (Pagel,
Edmunds, \& Smith 1980; Skillman 1989; Pilyugin 2000) and metal-rich
(Pagel et al. 1979; Edmunds \& Pagel 1984, McCall et al. 1985; Dopita
\& Evans 1986; Pilyugin 2001) \hii\ regions.

The traditional \rr23\ method typically results in metallicities 
within 0.2 dex of those calculated with the more accurate \telec\ abundances. 
Pilyugin (2000, 2001) improves on the standard \rr23\ method by
introducing a $P$ factor, which is calculated from the strong oxygen
lines. The $P$ factor replaces the temperature as a descriptor of the
conditions in the nebula and results in metallicities within 0.1 dex
of those calculated with \telec\ method.

We thus estimates the oxygen abundance of the two galaxies from the 
following Pilyugin's calibrations:

\begin{equation}
	\label{eq1}
	\displaystyle 12+\log(O/H)_{R_{23}}= 6.53 + 1.40\log(R_{23})
\end{equation}
\begin{equation}
	\label{eq2}
	12+\log(O/H)_{P}= \frac{R_{23}+54.2+59.45P+7.31P^{2}}{6.07+6.71P+0.371P^{2}+0.243R_{23}}
\end{equation}
\newline
where $P \equiv$ (\oiiia+\oiiib)/(\oiia+\oiib+\oiiia+\oiiib) is the 
excitation parameter. Equation~\ref{eq1} refers to the low-metallicity 
regime and equation~\ref{eq2} to the high-metallicity one. 
In order to compare our measurements with oxygen 
abundances reported in the literature for nearby and high-redshift 
star-forming galaxies, we also used McGaugh (1991) calibrations with 
the analytic expressions given in Kobulnicky et al. (1999). 
Oxygen abundances determined with these calibrations are listed in 
Table~\ref{tab4}. 
Since \halpha\ is detected in S2 and A2 but \hbeta\ is not, we 
adopt $I_{\rm H\beta}=I_{\rm H\alpha}/2.86$ assuming theoretical 
case B hydrogen recombination ratios for ionized gas with electron 
temperature of $10^4$ K (Osterbrock 1989). This approximation is 
very insensitive to the actual electron temperature but ignores 
effects of dust extinction. To evaluate this effect, we applied 
the extinction coefficients E(B-V)=0.3 and 0.4 for S2 and A2 respectively 
(see sect.~\ref{SFR}) and corrected emission-line fluxes. 
As expected, the correction for extinction raises \rr23\ by 
increasing the strength of \oii\ relative to \hbeta. The value of \doh\ 
is thus higher (by $0.2-0.3$ dex) on the low-metallicity branch and 
lower (by $0.1-0.3$ dex) on the high-metallicity branch, 
depending of the calibration used (see Table~\ref{tab4}). 

A complication with the ``strong lines'' method is that the dependence
of metallicity on \rr23\ is double valued. In the most metal-rich
\hii\ regions, \rr23\ is small because metals permit efficient
cooling, reducing the electronic temperature and the level of
collisional excitation. On the {\em upper}, metal-rich branch of the
relationship, \rr23\ increases as metallicity decreases via reduced
cooling and elevates the electronic temperature and the degree of
collisional excitation.  However, the relation between \rr23\ and O/H
changes when \doh\ $\sim$ 8.4 ($Z \sim 0.3$ \zsun).  As metallicity
decreases below \doh\ $\sim 8.2$, \rr23\ decreases once again. Even
though the reduced metal abundance further inhibits cooling and raises
the electron temperature on this {\em lower}, metal-poor branch, the
intensity of the oxygen emission lines drops because of the greatly
reduced oxygen abundance in the ionized gas. In this regime, the
ionization parameter also becomes important (e.g., McGaugh 1991).

Several methods have been proposed to break this degeneracy.  Contini
et al. (2002) showed that the best abundance indicators are the
\nii/\halpha\ and \nii/\oii\ line ratios which are virtually
independent of the ionization parameter. Both line ratios can thus be
used simultaneously to discriminate between the {\em lower} and the
{\em upper} branches in the O/H vs. \rr23\ relation. The abundance
indicators derived for S2 and A2 are listed in Table~\ref{tab4}. They
indicate clearly that S2 lies on the lower branch of the O/H
vs. \rr23\ relationship, whereas A2 lies on the upper
branch. Additional evidence for the low metallicity in S2 is given by
the strength of the C\,{\sc iv}$\lambda$1550 absorption line measured
in the rest-frame UV FORS1 spectrum of this object (Le Borgne et al.,
in preparation).  Using the relation between \doh\ and EW(C\,{\sc iv})
derived by Mehlert et al. (2002) using Heckman et al. (1998) data, we
deduce a metallicity $Z \sim$ 0.25 \zsun, which is compatible with the
value estimated using the strong optical line method.
 
\subsubsection{Nitrogen Abundance}

Nitrogen-to-oxygen abundance ratios (N/O) may be determined in the
absence of a measurement of the temperature-sensitive \oiiic\ emission
line using the algorithm proposed by Thurston et al.  (1996). Again,
this only requires the bright \nii, \oii\ and \oiii\ emission
lines. The relationship, which is based on the same premise as that
used in the relationship between the oxygen abundance and \rr23, is
calibrated using photoionization models.

First, an estimate of the temperature in the \ntwo\
emission region (\tnii) is given by the empirical
calibration between \tnii\ and \rr23\ (Thurston et al. 1996):
\begin{equation}
	\label{eq3}
t_{[\rm N\ II]}=6065+1600(\log R_{23})+1878(\log R_{23})^2+2803(\log R_{23})^3.
\end{equation}

The \ntwo\ temperature determined from the \rr23\ relation
can thus be used together with the observed strengths of \nii\
and \oii\ to determine the ionic abundance ratio N$^+$/O$^+$.
Pagel et al. (1992) gave the following formula based on
a five-level atom calculation:
\begin{equation}
\log \frac{{\rm N}^+}{{\rm O}^+}=\log \frac{\rm [N\, II]\lambda
6584}{\rm [O\, II]\lambda 3727}+0.307-0.02\log t_{\rm
[N\, II]}-\frac{0.726}{t_{\rm [N\, II]}},
\end{equation}
where the \ntwo\ temperature is expressed in units of $10^4$
K. Finally, we assume that N/O $\equiv$ N$^+$/O$^+$.  There has been
some discussion regarding the accuracy of this assumption (e.g.,
Vila-Costas \& Edmunds 1993) but Thurston et al. (1996) found that
this equivalence only introduces small uncertainties in deriving
N/O. The values of N/O derived for S2 and A2 are listed in
Table~\ref{tab4}, without extinction correction and assuming a
reddening E(B-V)=0.3 and 0.4 for S2 and A2 respectively. An upper
limit on N/O is given for S2 because the \nii\ line is not
detected. As expected, the extinction correction reduces (by $\sim 0.2$
dex) the N/O abundance ratio due to the increase of \oii\ relative to
\nii.

\subsection{Kinematics}
\label{kinematics}

We have measured the line-of-sight velocity dispersion $\sigma$ of the
ionized gas from the brightest emission lines (\oiii\ and \halpha) that are 
corrected for instrumental broadening. The values obtained are $55\pm6$ \kms\ 
for the brightest node of AC114-S2 and $105\pm15$ \kms\ for AC114-A2.
In order to derive masses for 
these objects and to compare with the typical values obtained for LBGs, we estimate 
the virial masses from the following equation:
\begin{equation}
	\label{eq10}
	\displaystyle M_{\rm vir} (M_{\sun})=1.16 \times 10^{6}  \sigma^2 r_{1/2}
\end{equation}
where $r_{1/2}$ is the half-light radius in kpc and $\sigma$ is the
velocity dispersion in \kms. This expression corresponds
to the ideal case of an homogeneous sphere with a uniform density
distribution. Table~\ref{tab6} summarizes the results.
In the present case, we have used deep HST-WFPC2 images of this field,
obtained in the red filter F702W, to compute the half-light radius,
after correcting for the tangential stretching induced by
gravitational lensing. In both cases, we have obtained $r_{1/2} \sim
0.2''$. 

\begin{table}[t]
\begin{center}
\caption{Velocity Dispersions and Virial Masses}
\label{tab6}
\begin{tabular}{c c c c c}
\hline \hline
Galaxy 	&$\sigma$ & $r_{1/2}^a$ & $r_{1/2}^b$ & $M_{\rm vir}$ \\
  & [\kms] & [\arcsec] &  [kpc] & [$10^{10}$ \msun] \\
\hline 
AC114-S2& $ 55\pm 6$ 		& 0.18	& 1.52	&$0.53\pm0.12$\\
AC114-A2& $105\pm15$ 		& 0.22	& 1.85	&$2.36\pm0.67$\\
\hline 	    
\end{tabular}	
\end{center}
$ ^a $ From HST WFPC2 images.\\
$ ^b $ $\Omega_{0} = 0.3$; $\Lambda = 0.7$; $H_{0} = 70$ \kmsmpc\ \\	
\end{table}

\begin{figure}[t]
\centering
\vspace{-1.2cm}
\includegraphics[width=7.0cm,angle=-90]{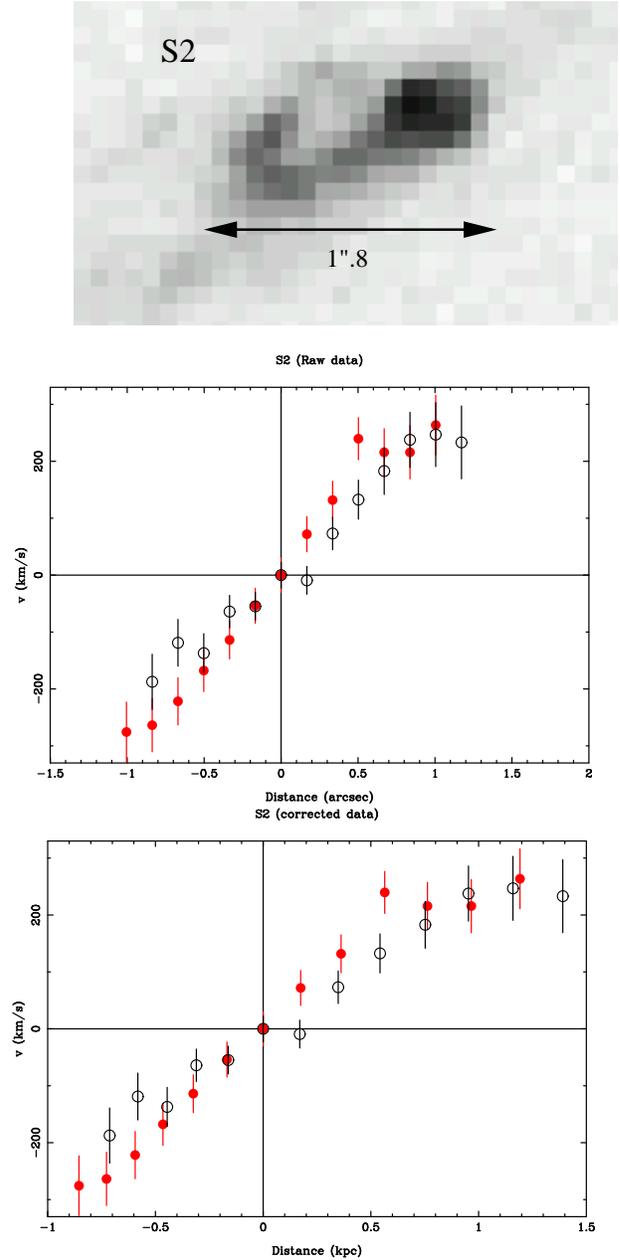}\\
\vspace{-1cm}
\includegraphics[width=6.0cm,angle=-90]{fig5b.ps}\\
\includegraphics[width=6.0cm,angle=-90]{fig5c.ps}
\caption{Spatially resolved emission lines in AC114-S2. 
{\it Top panel:} HST/WFPC2 ($R$-band/F702W) close image of the lensed 
galaxy AC114-S2. {\it Mid panel:} Raw velocity profiles (in \kms) versus 
angular distance (in arcsec), as measured on the 2D spectra from the 
central velocities of \halpha\ (open dots) and \oiiib\ (full dots) emission 
lines. {\it Bottom panel:} Lens corrected velocity profiles.}
\label{rotation}
\end{figure}

Thanks to the gravitational magnification, which strengths the 
image along the tangential direction, the line profiles of AC114-S2 are spatially 
resolved. Figure~\ref{rotation} displays the raw and lens-corrected velocity 
profiles measured from \halpha\ and \oiiib\ emission lines, after correction for 
the geometrical position of the object across the slit. 
The detailed lens model has been used 
to derive the tangential magnification along the object. The spatial resolution 
of the 2D spectra ranges between 0.5\arcsec\ and 1\arcsec\ without the tangential 
stretching induced by lensing (4 to 8 kpc with this cosmology). The velocity 
gradient seems to reach a ``plateau'' at $\sim \pm 1$ kpc from the center of the
object. In the hypothesis that this velocity gradient traces a rotation curve 
of $\pm 240$ ($\pm 30$) \kms\ in a disk-like geometry, the dynamical mass within 
the inner 1 kpc radius is $M_{\rm dyn}= R v^2 / G = 1.3 \pm 0.3 \times 10^{10}$ 
\msun. This value is a factor of 2 larger than the mass derived from the velocity 
dispersion.

The virial mass derived for AC114-A2, and the largest
determination for AC114-S2, a few $10^{10} M_{\sun}$, is of the
same order as the typical values found for LBGs (e.g. Pettini et
al. 2001). In contrast, AC114-S2 is about one magnitude brighter 
than AC114-A2, and one magnitude fainter than the typical LBGs in the 
rest-frame $B$-band. Thus, the $M/L_{\rm B}$ ratio of AC114-S2 in 
solar units ranges between 0.20 and 0.5, the lower value being 
quite similar to that of the LBGs at $z \sim 3$, whereas AC114-A2 has a
$M/L_{\rm B} = 2$, a ratio $\sim 20$ times higher than for LBGs. After 
dust extinction correction, we obtain $M/L_{\rm B} \sim$ 0.15 and 0.6
for AC114-S2 and AC114-A2 respectively, using the dynamical mass for S2.
It is probable that both velocities and masses have been underestimated 
because of the limited sensitivity of the observations, since we see 
only the inner cores of the galaxies where star formation is the strongest.

\section{Discussion}
\label{discussion}

\begin{table*}[t]
\begin{center}
\caption{Physical properties of high-redshift ($z \geq 2$) galaxies. 
Abundance ratios and SFRs are without extinction correction.}
\label{tab5}
\begin{tabular}{l c c c c c c c}
\hline\hline
Galaxy 			     & $z$  &$M_{\rm B}$ &12+log(O/H)& log(N/O)		&${\rm SFR_{UV}}$&${\rm SFR_{H\alpha}}$&$M_{\rm vir}$ ($10^{10}$ \msun)\\
\hline  						    
AC 114-S2\dotfill	     &1.867 &$-$20.60	 &$7.25\pm 0.2$ & $<-1.32$	& 4.4	&30		   &$0.53\pm0.12$\\
AC 114-A2\dotfill 	     &1.869 &$-$19.80    &$8.94\pm 0.2$ & $-0.42\pm0.2$	& 1.1	&15		   &$2.36\pm0.67$\\
\hline				       		     
CFg \dotfill		     &2.313 &$-$22.28    &$7.8-8.7$     & ...		&11 	&54		   &7.0\\
MS 1512-cB58 \dotfill 	     &2.729 &$-$22.04	 &$8.39\pm 0.2$ & $-1.24\pm0.2$	&16	&18		   &1.8 \\
Lynx 2-9691 \dotfill 	     &2.888 &$-$23.19	 &$8.3-8.8$	& ...		&11 	&80		   &5.2 \\
Q0201+113 C6 \dotfill 	     &3.055 &$-$22.20    &$7.6-8.8$     & ... 		&27	&17		   &1.2 \\ 
SSA22a D3 \dotfill 	     &3.069 &$-$22.59	 &$8.0-8.6$     & ... 		&38	&24		   & ... \\ 
Q1422+231 D81 \dotfill 	     &3.104 &$-$22.91	 &$7.7-8.7$     & ... 		&45	&74		   & ... \\
DSF 2237+116 C2 \dotfill     &3.333 &$-$23.97	 &$7.6-8.8$     & ... 		&44	&75	           &5.5 \\
B2 0902+342 C12 \dotfill     &3.387 &$-$22.84	 &$7.6-8.8$     & ... 		&42	&61		   & ... \\
\hline 	    
\end{tabular}	
\end{center}
\end{table*}

Table~\ref{tab5} summarizes the physical properties for high-redshift
($z\ga 2$) galaxies. To date there is only one LBG in which chemical
abundances have been determined with some degree of confidence, the
gravitationally lensed galaxy MS 1512-cB58 at $z=2.729$ (Teplitz et
al. 2000; Pettini et al. 2002a). Here we provide accurate chemical
abundance measurements for the two high-redshift galaxies: AC114-S2
and AC114-A2.

In terms of SFRs without extinction correction, the two lensed low-luminosity 
galaxies A2 and S2 have ${\rm SFR_{1500} \ll SFR_{H\alpha}}$, while in LBGs at 
$z \sim 3$ ${\rm SFR_{1500}} \sim {\rm SFR_{H\alpha}}$ (within 15\%). 
Note that a factor of two still remains between the two SFRs of S2 when 
using the latest calibrations by Rosa-Gonzalez et al. (2002). As shown in 
Section~\ref{SFR}, a large E(B-V) value is needed to reconcile the UV/optical 
SFR values for these two objects. Unfortunately, we cannot derive an independent 
reddening estimate for S2 and A2, e.g. using the Balmer decrement
\halpha/\hbeta.. 

\subsection{The metallicity--luminosity relationship}

\begin{figure}[t]
\includegraphics[clip=,angle=0,width=8.8cm]{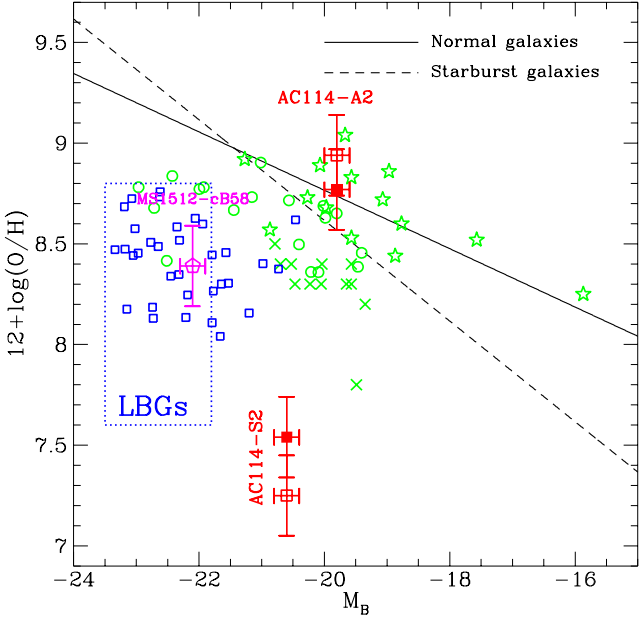}
\caption{Oxygen abundance (\doh) as a function of $B$-band absolute magnitude.
The metallicity--luminosity relations for nearby ``normal'' (solid line; 
Kobulnicky \& Zaritsky 1999) and starburst (dashed line; Mouhcine \& Contini 2002; 
Melbourne \& Salzer 2002) galaxies are shown. The location of the two $z \sim 1.9$ galaxies 
S2 and A2 in the lensing cluster AC 114 is shown without extinction correction (empty 
squares) and assuming a reddening E(B-V)=0.3 and 0.4 for S2 and A2 respectively 
(filled squares). We show for comparison samples of intermediate-redshift galaxies: 
UV-selected galaxies at $z \sim 0.1-0.4$ (circles; Contini et al. 2002), emission-line 
galaxies at $z \sim 0.1-0.5$ (stars; Kobulnicky \& Zaritsky 1999), and at 
$z \sim 0.5-0.7$ (crosses; Hammer et al. 2001). The location of high-redshift 
($z \sim 3$) Lyman break galaxies (LBGs) is shown as a blue box 
encompassing the range of O/H and \mabs\ derived for these objects (Pettini et al. 2001). 
High-redshift ($1.4<z<3.4$) galaxies with oxygen abundances and $B$-band magnitudes 
derived from rest-frame UV spectra (Mehlert et al. 2002) are shown as blue squares.}
\label{oh_vs_mb}
\end{figure}

\begin{figure}[t]
\includegraphics[clip=,angle=0,width=8.8cm]{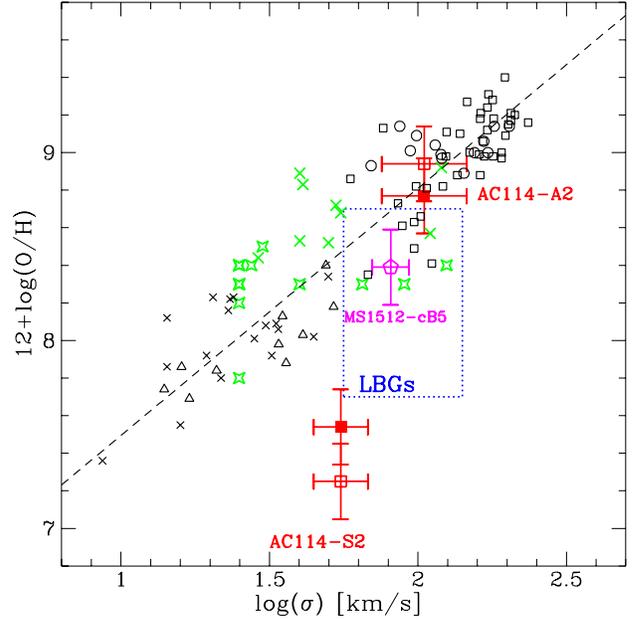}
\caption{Oxygen abundance (\doh) as a function of the
line-of-sight velocity dispersion ($\sigma$).
The location of AC114-S2 and AC114-A2 is shown without extinction correction (empty 
squares) and assuming a reddening E(B-V)=0.3 and 0.4 for S2 and A2 respectively 
(filled squares). We show for comparison samples of nearby galaxies: \hii\ galaxies 
(triangles; Telles \& Terlevich 1997), dwarf galaxies (crosses; Richer \& McCall 1995), 
late-type spirals (squares; Zaritsky, Kennicutt \& Huchra 1994) and early-type spirals 
(circles; Oey \& Kennicutt 1993). The dashed line is a linear fit to these samples. 
Intermediate-redshift galaxy samples are: 
emission-line galaxies at $z \sim 0.1-0.5$ (stars; Kobulnicky \& Zaritsky 1999), and at 
$z \sim 0.5-0.7$ (crosses; Hammer et al. 2001). The location of high-redshift ($z \sim 3$) 
Lyman break galaxies (LBGs) is shown as a blue box encompassing the range of O/H and 
$\sigma$ derived for these objects (Pettini et al. 2001).}
\label{oh_vs_sig}
\end{figure}

We can study the evolution with redshift of the fundamental scaling relation 
between galaxy luminosity and metallicity (Contini et al. 2002 and references therein). 
For nearby galaxies, this relation extends over $\sim 10$ magnitudes in 
luminosity and $\sim 2$ dex in metallicity.

The high-redshift sample is best compared 
to nearby galaxies where metallicities are derived using the same empirical strong 
line method. Figure~\ref{oh_vs_mb} shows the relations for nearby i) ``normal'' 
irregular and spiral galaxies (solid line; Kobulnicky \& Zaritsky 1999) and ii) 
starburst galaxies (dashed line; Mouhcine \& Contini 2002; Melbourne \& Salzer 
2002), as well as the high-redshift sample from Table~\ref{tab5}. 
Three samples of intermediate-redshift galaxies are also shown for comparison: 
UV-selected star-forming galaxies at $z\sim 0.1-0.4$, emission-line galaxies 
at $z\sim 0.1-0.5$ (Kobulnicky \& Zaritsky 1999), and luminous compact emission-line 
galaxies at $z\sim 0.5-0.7$ (Hammer et al. 2001). 

It is immediately obvious that high-redshift galaxies do not conform
to today's luminosity--metallicity relation for both ``normal'' and
starburst galaxies. Even allowing for the uncertainties in the
determination of O/H, high-redshift galaxies have much lower oxygen
abundances than one would expect from their luminosities. This result,
already revealed by previous studies (Kobulnicky \& Koo 2000; Pettini
et al. 2001; Contini et al. 2002), is secured with the addition of the
low-luminosity and low-metallicity galaxy S2.

One interpretation of this result is that high-redshift galaxies
are undergoing strong bursts of star formation which raise their
luminosities above those of nearby galaxies with similar chemical
composition.  Another possibility is that the whole
metallicity--luminosity relation is displaced to lower abundances at
high redshifts, when the Universe was younger and the time available
for the accumulation of the products of stellar nucleosynthesis was
shorter. It should be possible to quantify this effect by measuring
metallicities in samples of galaxies at different redshifts
(e.g. Mehlert et al. 2002). As shown in Figure~\ref{oh_vs_mb},
intermediate-redshift galaxies seem to follow the
luminosity--metallicity relation derived for nearby starburst galaxies
(dashed line). Unfortunately, to date there is no sample of galaxies
with known chemical abundances to fill the gap between $z\sim 0.5$ and
$z\sim 2$.

The location of A2 is more surprising. It does not follow the trend of
high-redshift objects. Instead, it lies on the luminosity--metallicity
relation of nearby objects and has the highest metallicity of any
high-redshift ($z \geq 2$) galaxy.

We compute the intrinsic $M_{\rm I}$ from the observed $K$-band magnitudes 
(2.2 \micron\ corresponds to 7600 \AA\ at $z \sim 1.9$) for S2 and A2. 
We found the same value $M_{\rm I}=-23.2$ for S2 and A2. 
This suggests that S2 and A2 may have the same stellar masses, since in the $I$ band 
we have access to the old stellar population which dominates the stellar mass. 
But the relatively high nitrogen abundance in AC114-A2 and MS 1512-cB58 suggests
the presence of an older population of stars than in AC114-S2. 

In Figure~\ref{oh_vs_sig} we plot the oxygen abundance (\doh) against
the line-of-sight velocity dispersion ($\sigma$) for S2 and A2 and
other relevant samples of nearby (Telles \& Terlevich 1997; Richer \&
McCall 1995; Zaritsky, Kennicutt \& Huchra 1994; Oey \& Kennicutt
1993), intermediate (Kobulnicky \& Zaritsky 1999; Hammer et al. 2001)
and high redshift (see Table~\ref{tab5}) galaxies.  This diagram
corresponds to a crude mass--metallicity sequence.  The location of
AC114-A2 is fully compatible with that of low-redshift galaxies.
However, the oxygen abundance of AC114-S2 is much smaller than the
corresponding value for low-$z$ galaxies of similar velocity
dispersions. The situation is even more dramatic if we 
consider the velocity gradient derived under the disk-like 
rotation curve hypothesis (see sect.~\ref{kinematics}).
Taken at face value, it would appear that the slope of
the mass--metallicity relation for galaxies with $z \geq 2$ is
different from that of low-$z$ galaxies.

\subsection{The N/O versus O/H relationship}

\begin{figure}
\includegraphics[clip=,angle=0,width=8.8cm]{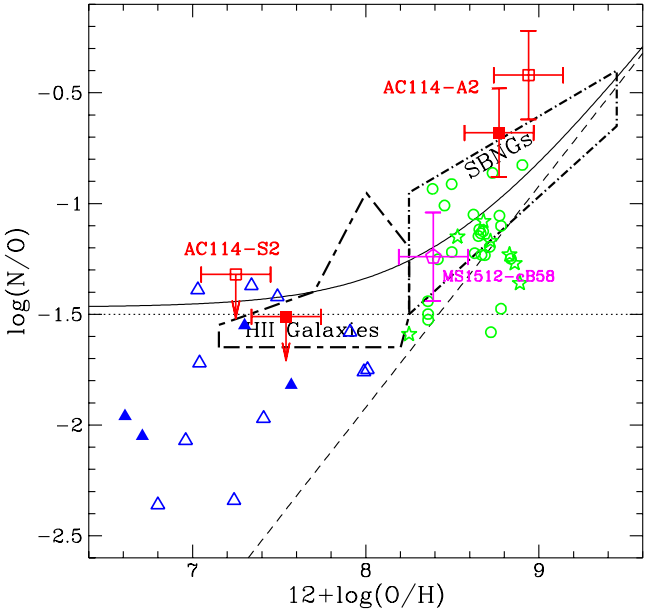}
\caption{Nitrogen-to-oxygen (N/O) abundance ratio as a function
of oxygen abundance (\doh). The location of the two $z \sim 1.9$ 
galaxies S2 and A2 in the lensing cluster AC 114 is shown without 
extinction correction (empty squares) and assuming a reddening 
E(B-V)=0.3 and 0.4 for S2 and A2 respectively (filled squares). 
Two comparison samples of nearby star-forming 
galaxies are shown (see Contini et al. 2002 for references): 
starburst nucleus galaxies (dot -- short dash line) and \hii\ galaxies 
(short dash -- long dash line). We show, for comparison, samples of 
intermediate-redshift galaxies: UV-selected galaxies at 
$z \sim 0.1-0.4$ (circles; Contini et al. 2002), and emission-line 
galaxies at $z \sim 0.1-0.5$ (stars; Kobulnicky \& Zaritsky 1999). 
The location of Damped Ly$\alpha$ systems (triangles, filled symbols 
are upper limits; Pettini et al. 2002b) is shown for comparison. 
Theoretical curves for a {\em primary} (dotted line), a 
{\em secondary} (dashed line), and a {\em primary + secondary} 
(solid line) production of nitrogen (Vila-Costas \& Edmunds 
1993) are shown.}
\label{no_vs_oh}
\end{figure}

In Figure~\ref{no_vs_oh}, we examine the location of S2 and A2 
in the N/O versus O/H relationship. The behavior of N/O with
increasing metallicity offers clues about the history of chemical evolution
of galaxies and the stellar populations responsible
for producing oxygen and nitrogen. The location of the two $z \sim 1.9$ 
galaxies AC114-S2 and AC114-A2 are shown without 
extinction correction (empty squares) and assuming a reddening 
E(B-V)=0.3 and 0.4 for S2 and A2 respectively (filled squares). 
For comparison, we show in Figure~\ref{no_vs_oh} the location of 
nearby star-forming galaxies: starburst nucleus galaxies (SBNGs) and 
\hii\ galaxies (see Contini et al. 2002 for references). We also plot 
samples of intermediate-redshift galaxies: UV-selected galaxies 
at $z \sim 0.1-0.4$ (Contini et al. 2002), and emission-line 
galaxies at $z \sim 0.1-0.5$ (Kobulnicky \& Zaritsky 1999). 
The location of Damped Ly$\alpha$ systems (Pettini et al. 2002b) 
is also shown.
 
The sample of high-redshift ($z \geq 2$) galaxies with measured N/O
abundances is still very small: A2 and S2 from this paper, and MS
1512-cB58 at $z \sim 2.7$ (see Table~\ref{tab5}). Surprisingly, these
three galaxies have very different locations in the N/O vs. O/H
diagram. S2 is a low-metallicity object (\doh\ $\sim 7.25$; Z $\sim
0.03$ \zsun) with a low N/O ratio (N/O $< -1.32$), similar to those
derived in the most metal-poor nearby \hii\ galaxies. In contrast, A2
is a metal-rich galaxy (\doh\ $\sim 8.94$; Z $\sim 1.3$ \zsun) with a
high N/O abundance ratio (N/O $\sim -0.42$), similar to those derived
in the most metal-rich massive SBNGs. The position of MS 1512-cB58 is
intermediate between these two extremes showing abundance ratios
typical of low-mass SBNGs and intermediate-redshift galaxies (see
Fig.~\ref{no_vs_oh}).

A natural explanation for the variation of N/O at constant metallicity
is the time delay between the release of oxygen and that of nitrogen
into the ISM (e.g. Contini et al. 2002, and references therein), while
maintaining a universal IMF and standard stellar nucleosynthesis. The
``delayed-release'' model assumes that star formation is an
intermittent process in galaxies (e.g. Edmunds \& Pagel 1978; Garnett
1990; Coziol et al. 1999) and predicts that the dispersion in N/O is
due to the delayed release of nitrogen produced in low-mass
longer-lived stars, compared to oxygen produced in massive,
short-lived stars.

Following this hypothesis and new chemical evolution models (Mouhcine \& 
Contini 2002), we might interpret the location of S2 and A2 in the N/O 
versus O/H diagram in terms of star formation history and evolutionary stage 
of these galaxies. The low O/H and N/O abundance ratios found in S2 
might indicate a relatively young age for this object which experienced 
two or three bursts of star formation at most in the recent past. A2 seems 
on the contrary much more evolved. The location of this galaxy in the N/O 
versus O/H diagram indicates a relatively long star formation history with 
numerous powerful and extended starbursts.  

\section{Conclusions}
\label{conclusions}

We have presented in this paper the firsts results of our
NIR spectroscopic survey of highly magnified high-redshift galaxies in the
core of lensing clusters. Our first targets were AC114-S2 and
AC114-A2, two lensed sources at $z \sim 1.9$. We have obtained, for 
the first time, the rest-frame optical emission lines 
of galaxies 1-2 magnitudes fainter than the LBGs 
studied in field surveys (e.g. Pettini et al. 1998, 2001; Kobulnicky 
\& Koo 2000). The main results obtained in this paper are the following:

\begin{enumerate}

\item The SFRs derived from the \halpha\ emission-line are systematically
higher than the values deduced from the 1500\AA\  flux
without dust extinction correction. The reddening
values needed to obtain SFR$_{\rm H\alpha}$/SFR$_{\rm UV}$ $\sim 1$ 
using the Calzetti et al. (2000) reddening law are E(B-V)=0.29 for 
AC114-S2 and E(B-V)=0.40 for AC114-A2, leading to corrected SFRs 
of 73 \msunyr\ and 51 \msunyr\ respectively. These results suggest that 
large dust extinction corrections are needed for intrinsically faint 
galaxies compared to field (brighter) LBG samples. 

\item The large spectral coverage obtained, from \oii\ to
\halpha+\nii, allows us to set strong constraints on the
chemical abundance ratios (O/H and N/O) of these two faint galaxies. 
The behavior of S2 and A2 in terms of metallicity is different, 
and the two objects are also different from typical LBGs at $z \sim 3$, 
suggesting a different star formation history for galaxies of different 
luminosities. 

\item The virial masses derived from the line-of-sight velocity dispersions 
are of the order of $5 \times 10^{9}$ \msun\ for AC114-S2 and 
$\sim 10^{10}$ \msun\ for AC114-A2. The line profiles of AC114-S2 are 
spatially resolved, leading to a velocity gradient of $\pm 240$ \kms, which 
yields a dynamical mass of $1.3 \pm 0.3 \times 10^{10}$ \msun\ within the inner 
1 kpc radius, under the hypothesis of a rotation curve with disk-like geometry.
Although the $M/L_{\rm B}$ ratio of AC114-S2 is
quite similar, within a factor of 2, to the values found for 
field ($\sim$ 1 mag brighter) LBGs, the corresponding value for 
AC114-A2 is about one order of magnitude higher. After 
dust extinction correction, we obtain $M/L_{\rm B} \sim $ 0.15 and 0.6
for S2 and A2 respectively.

\end{enumerate}

In terms of SFRs, mass-to-light and chemical abundance ratios, 
the behaviour of S2 and A2 is rather close to that described by 
Guzm\'an et al. (1997) for their sample of compact emission-line 
galaxies in the HDF at $0.7 \la z \la 1.4$. S2 would correspond to 
a young star-forming \hii\ galaxy, whereas A2 is more likely an 
evolved starburst.

The results obtained on the physical properties of AC114-S2 and
AC114-A2 suggest that high-$z$ objects of different luminosities could
have quite different star formation histories. However, the number of well
observed high redshift objects is currently very small and larger samples
in both redshift (1.5 $\ltapprox$ z $\ltapprox$ 6) and luminosity are
required. This could be achieved with the new generation of
multi-object NIR spectrographs for the 10m class telescopes, such as
KMOS on the VLT or EMIR on the GTC.

\acknowledgements

We are grateful to G. Bruzual, F. Courbin, G. Golse, H.~A. Kobulnicky, 
M. Pettini, D. Schaerer, J. Richard and J. Gallego for useful discussions 
on this particular program. 
Part of this work was supported by the
French {\it Conseil R\'egional de la Martinique},
by the French {\it Centre National de la Recherche Scientifique}, the TMR {\it
Lensnet} ERBFMRXCT97-0172 (http://www.ast.cam.ac.uk/IoA/lensnet) and
the ECOS SUD Program.

\end{document}